\documentclass{ws-procs9x6}

\begin{document}

\title{Determination of the low $Q^2$ evolution of the Bjorken integral}

\author{A. Deur}

\address{Thomas Jefferson National Accelerator Facility \\
12000 Jefferson Avenue, Newport News, VA23606, USA\\ 
E-mail: deurpam@jlab.gov}

\maketitle

\abstracts{
We report on an experimental determination of the 
$Q^2$-dependence of the Bjorken sum using data from
Jefferson Lab Hall A and Hall B in the range $0.16<Q^2<1.1$ GeV$^2$. 
A twist analysis is performed. Overall, the higher twist corrections are
found to be small due to a cancellation between the twist 4
and 6 terms.}

\section{The GDH and Bjorken Sum Rules}

A main reason to study the generalized GDH sum is to understand the
transition from the hadronic to the partonic descriptions of the nucleon.
The generalized GDH sum is in principle calculable at
any $Q^2$, which makes it an ideal tool to study the hadronic to partonic 
transition. This topic was covered in the symposium\cite{Choi}. 
However, the validity domains of the available chiral 
perturbation theory ($\chi$PT) and pQCD calculations do not overlap.
Lattice QCD should provide the bridge between the two domains but no 
calculation is available yet.

The Bjorken sum rule\cite{Bjorken} has been a cornerstone of polarized pQCD 
studies. At leading twist, it reads:
\begin{equation}
\int_{0}^{1^{-}}(g_{1}^{p}-g_{1}^{n})dx=\frac{g_{a}}{6}
[ 1-\frac{\alpha_{\rm{s}}}{\pi}-3.58\left(\frac{\alpha_{\rm{s}}}{\pi}\right)^{2}-20.21\left(\frac{\alpha_{\rm{s}}}{\pi}\right)^{3} +...]
\label{eqn:bj}
\end{equation}
\noindent 
where $g_a$ is the nucleon axial charge. The connection between a
generalized GDH sum and the Bjorken sum was made by M. Anselmino \emph{et al.} 
\cite{Anselmino}. More recently, X. Ji and J. Osborne made the 
reason for the connection clear: the GDH and the Bjorken 
sum rules are two particular cases of a more general sum rule\cite{Ji}. 
The extended GDH sum rule, as generalized in Ref.\cite{Ji}, links the first 
moment of the spin structure functions to the spin dependent forward Compton
scattering amplitudes. Hence, the relation between the generalized GDH and 
Bjorken sums is:
\begin{equation}
\int_{0}^{1^{-}}g_{1}^{p}-g_{1}^{n}dx=
\frac{Q^2}{16\pi^2\alpha}(\rm{GDH}^p(Q^2)-\rm{GDH}^n(Q^2))
\label{eqn:link}
\end{equation}
Considering this $p-n$ flavor non-singlet quantity yields many advantages. 
\begin{enumerate}

\item {At large $Q^2$, we have a sum rule (the Bjorken sum rule) without 
hypothesis beyond QCD (as opposed to the Ellis-Jaffe sum rule\cite{EJ}
for singlet quantities).}
\item { The estimations of the unmeasured low-$x$ part of the integral are
more reliable.}
\item {The pQCD evolution equations are simpler. }
\item {At low $Q^2$, the $\chi$PT calculations are more reliable due 
to the cancellation of the $\Delta_{1232}$ contribution. }

\end{enumerate}

These advantages might help in extending
the validity domains  of pQCD and $\chi$PT. It is conceivable that the 
hadron-parton gap 
might be bridged, allowing one to fundamentally describe 
the nucleon structure at all scales\cite{Volker} for the first time. 
Hence the Bjorken sum
is arguably one of the most convenient quantities to measure in the
resonance region to understand the hadron-parton transition. \\

\begin{figure}[ht]
\centerline{\epsfxsize=3.8in\epsfbox{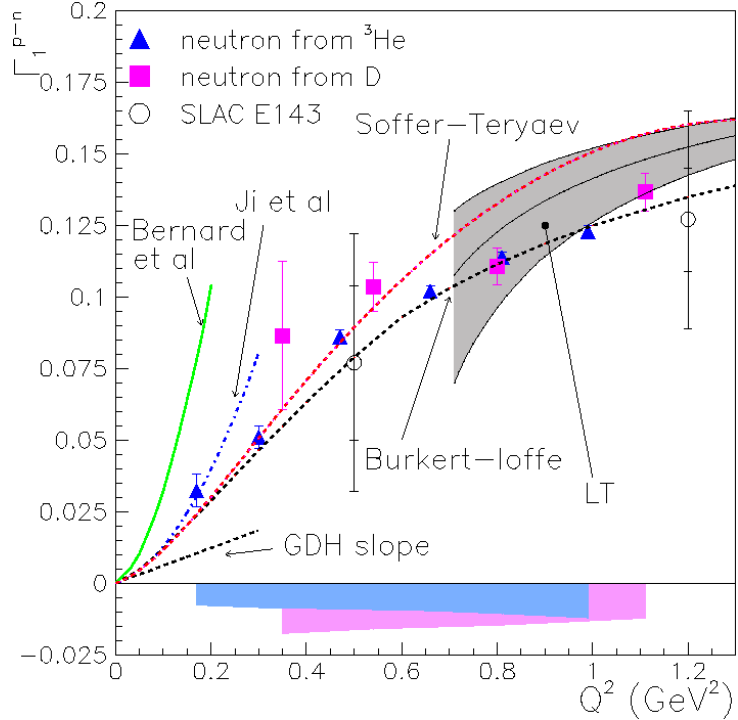}}   
\caption{$Q^2$-evolution of the Bjorken sum. The Bjorken sum formed
using neutron data extracted from $^3$He (D) data is shown by the triangles 
(squares) and the horizontal band spanning from $Q^2=0.16$ to 1.0 GeV$^2$ 
(from $Q^2=0.35$ to 1.1 GeV$^2$) is the corresponding systematic uncertainty.  
The resonance data from SLAC E143 are also shown (circles). The pQCD 
calculation at leading twist is represented by the gray band. Two $\chi$PT 
calculations (Ji \emph{et al.} and Bernard  \emph{et al.}) can be seen at low 
$Q^2$ as well as the GDH slope which constrains the Bjorken sum near the 
photon point. The dashed curves are the predictions of two phenomenological
models (Burkert and Ioffe, bottom curve, and Soffer and Teryaev, top curve).  
\label{bj}}
\end{figure}

Precise data are available from the Thomas Jefferson National
accelerator facility (JLab). Results were published on the 
proton\cite{eg1a proton} and deuterium\cite{eg1a deuteron}, from CLAS in 
Hall B 
and on $^3$He from Hall A\cite{E94010-1,E94010-2}. We used these data
to extract the Bjorken sum from $Q^2=0.16$ to 1.1 GeV$^2$. 
To combine proton and neutron 
data, the $^3$He data were reanalyzed at the same $Q^2$ as those of 
Ref.\cite{eg1a proton}. For consistency, the unmeasured low-$x$ part of 
the integral was 
re-evaluated for the three data sets using a consistent prescription\cite{BT}. 
The part beyond the validity range of Ref.\cite{BT} was
estimated using a Regge form and forcing the total integral as measured by
the SLAC E155 experiment\cite{E155} at $Q^2$=5 GeV$^2$ and completed by the 
estimation\cite{BT} and our Regge form, to verify the Bjorken sum rule at 
$Q^2$=5 GeV$^2$. The results are shown on Figure~\ref{bj}. The 
elastic contribution is not included. The negative horizontal bands give 
the systematic uncertainties on the two data sets. Also plotted are the
SLAC E143 results in the resonance region\cite{E143}. Two $\chi$PT 
calculations, from Bernard \emph{et al.}\cite{meissnerchipt} and 
Ji \emph{et al.}\cite{jichipt}, are shown at low $Q^2$. At $Q^2=0$, the 
Bjorken sum is constrained by the GDH sum rule (see eq. \ref{eqn:link}). 
At higher $Q^2$ the leading twist calculation up to third order in $\alpha_s$ 
is shown by the gray band. Its width is due to the uncertainty on $\alpha_s$. 
The model of Soffer and Teryaev\cite{ST} overestimates
the data at large $Q^2$. An improved version accounting for the
pQCD radiations was presented\cite{Soffer} and seems to agree
better with the data. The calculation from Burkert and Ioffe\cite{AO}
agrees well with the data. 

In the particular case of the$\chi$PT calculation done in the heavy Baryon 
approximation\cite{jichipt}, the comparison with the data may
indicate that the $\chi$PT domain of validity is indeed extended
since they agree, up to about $Q^2=0.25$
GeV$^2$ (to be compared to $Q^2=0.1$ GeV$^2$  typically for singlet
quantities). However, the calculations from Bernard 
\emph{et al.}\cite{meissnerchipt}, that do not employ the heavy Baryon 
approximation, do not support this conclusion.
In any case, a gap between the  $\chi$PT calculations and the pQCD calculation
clearly remains. More details can be found in Ref.~\cite{bj} 

\section{Twist Analysis}

It is remarkable that the data agree with the leading 
twist calculation down to quite low $Q^2$. This may indicate that higher
twist terms are small or cancel each other. To
quantitatively address this, we performed a twist analysis.
The coefficient $\mu^{p-n}_4$ of the $1/Q^2$ correction to Eq. \ref{eqn:bj} is:
\begin{eqnarray}
\mu_4^{p-n}
&=& \frac{M^2}{9}
    \left( a^{p-n}_2 + 4 d^{p-n}_2 + 4 f^{p-n}_2 \right),
\end {eqnarray}
 where $a_2^{p-n}$ is the target mass correction given by the second
moment of $g_1^{p-n}$, and $d_2^{p-n}$ is a twist-3 matrix element given by
\begin{eqnarray}
d^{p-n}_2
&=& \int_0^1 dx~x^2 \left( 2g^{p-n}_1 + 3g^{p-n}_2 \right).
\end{eqnarray}
The term $a_2^{p-n}$ is computed using our data and 
$d_2^{p-n}$ is obtained from SLAC and JLab data\cite{a1n}.
The twist 4 term $f_2^{p-n}$ is extracted from a fit of the Bjorken sum
including the elastic contribution. After re-estimating
the low-x part of the world data using our same 
method, we fit our data together with the world data in the 0.66-15 
GeV$^2$ $Q^2$-range. To verify that the twist series is convergent, 
the next twist term $\mu^{p-n}_6$ was included. We obtain, for
$Q^2=1$ GeV$^2$, 
$f_2^{p-n}=-0.17 \pm 0.05 (uncor) ^{+0.04}_{-0.05}(cor)$ and 
$\mu^{p-n}_6=0.09 \pm 0.02 (uncor) \pm 0.01 (cor)$ GeV$^4$ where 
$uncor$ ($cor$) specifies the uncertainties due 
to the point to point uncorrelated ($correlated$) uncertainty on the 
JLab data. Comparing
$\mu^{p-n}_6/Q^4=0.09 \pm 0.02$ to $\mu^{p-n}_4/Q^2 \simeq -0.06 
\pm 0.02$, we find that the $Q^{-2}$ and $Q^{-4}$ terms have opposite sign 
and similar magnitude, making the overall twist correction small at 
$Q^2=1$ GeV$^2$. This result may explain why many experimental
JLab results tend to indicate that pQCD works down to surprisingly
low $Q^2$. 

\section{Summary and outlook}

We have extracted the Bjorken sum in the $Q^2$ range of 0.16-1.1 GeV$^2$. 
Compared to singlet quantities, the $\chi$PT calculation seems to agree with 
the data over a larger $Q^2$ range only in the case of the Heavy Baryon 
approximation. Such possible improvement is  not seen with the calculations of 
Ref.\cite{meissnerchipt}. This last point is not unlike the conclusion
reached in Ref.\cite{jpchen} where $\chi$PT calculations were compared to 
measurements of the generalized spin polarizability $\delta_{LT}$, a quantity 
in which the $\Delta$ degrees of freedom are also suppressed. 
At any rate the parton to hadron gap, if smaller, is not bridged yet.
The magnitudes of the higher twists were extracted, in particular $f_2^{p-n}$. 
The higher twist effects appear to be small, due to a cancellation of the
1/Q$^2$ and 1/$Q^4$ terms. The analysis of new proton and deuterium data from 
CLAS will be finalized soon. These cover a larger $Q^2$ range 
(0.05 to 4 GeV$^2$) with improved statistics \cite{Dodge}. Data at even lower 
$Q^2$ are expected to be available in the upcoming years from both Jefferson 
Lab Hall A ($^3$He and neutron\cite{E97110}) and Hall B (proton)\cite{E03006}.

\end{document}